\documentclass[10pt,conference,a4paper]{IEEEtran}
\usepackage[T1]{fontenc}
\usepackage{amsmath}
\usepackage[cmintegrals]{newtxmath}
\usepackage{url}
\usepackage{soul,color,graphicx,float,epstopdf}
\usepackage{tablefootnote,textcomp,gensymb}
\usepackage{eurosym,booktabs,multirow}
\usepackage[caption=false]{subfig}
\usepackage{amsfonts} 
\usepackage{algorithmic}
\usepackage{array}
\usepackage{stfloats}
\usepackage{float}
\usepackage{mathtools}
\usepackage{graphicx}
\usepackage{pdfpages}
\usepackage{subfig} 
\usepackage{balance}
\usepackage{xcolor}
\usepackage{comment}
\usepackage[normalem]{ulem}
\usepackage{dsfont}
\usepackage[linesnumbered, ruled]{algorithm2e}
\usepackage{enumitem}
\usepackage{tikz}
\usetikzlibrary{arrows,shapes,positioning,shadows,trees}

\tikzset{
  basic/.style  = {draw, text width=4cm, drop shadow, font=\sffamily, rectangle},
  root/.style   = {basic, rounded corners=2pt, thin, align=center,
                   fill=red!20},
  level 2/.style = {basic, rounded corners=6pt, thin,align=center, fill=blue!20,
                   text width=9em},
  level 3/.style = {basic, thin, align=left, fill=pink!20, text width=8em}
}

\definecolor{Orange}{rgb}{1,0.5,0}
\newcommand{\eqr}[1]{$#1$}

\newcommand*{\rom}[1]{\expandafter\@slowromancap\romannumeral #1@}

\newcommand{\todo}[1]{\textsf{\textbf{\textcolor{Orange}{[#1]}}}}

\newcommand{\PZ}[1]{\textcolor{black}{#1}}

\begin{document}
\title{Transmit Power Control for Indoor Small Cells: A Method Based on Federated Reinforcement Learning}
\author{\IEEEauthorblockN{
Peizheng Li\IEEEauthorrefmark{1},
Hakan Erdol\IEEEauthorrefmark{1},
Keith Briggs\IEEEauthorrefmark{2},
Xiaoyang Wang\IEEEauthorrefmark{1},
Robert Piechocki\IEEEauthorrefmark{1},\\
Abdelrahim Ahmad\IEEEauthorrefmark{3},
Rui Inacio\IEEEauthorrefmark{3},
Shipra Kapoor \IEEEauthorrefmark{2},
Angela Doufexi\IEEEauthorrefmark{1},
Arjun Parekh \IEEEauthorrefmark{2}
}\\ 
\IEEEauthorblockA{\IEEEauthorrefmark{1} University of Bristol, UK;
\IEEEauthorrefmark{2} Applied  Research, BT, UK; \IEEEauthorrefmark{3} Vilicom UK Ltd.\\
Email: {\{peizheng.li, hakan.erdol, xiaoyang.wang, A.Doufexi,  R.J.Piechocki\}@bristol.ac.uk}\\
{\{keith.briggs, shipra.kapoor, arjun.parekh\}@bt.com}; 
{\{Abdelrahim.Ahmad, Rui.Inacio\}@vilicom.com}
}}
\maketitle

\begin{abstract}
Setting the transmit power setting of 5G cells has been a long-term topic of discussion, as optimized power settings can help reduce interference and improve the quality of service to users. Recently, machine learning (ML)-based, especially reinforcement learning (RL)-based control methods have received much attention. However, there is little discussion about the generalisation ability of the trained RL models. This paper points out that an RL agent trained in a specific indoor environment is room-dependent, and cannot directly serve new heterogeneous environments. Therefore, in the context of \PZ{Open Radio Access Network (O-RAN)}, this paper proposes a distributed cell power-control scheme based on Federated Reinforcement Learning (FRL). Models in different indoor environments are aggregated to the global model during the training process, and then the central server broadcasts the updated model back to each client. The model will also be used as the base model for adaptive training in the new environment. The simulation results show that the FRL model has similar performance to a single RL agent, and both are better than the random power allocation method and exhaustive search method. The results of the generalisation test show that using the FRL model as the base model improves the convergence speed of the model in the new environment.
\end{abstract}


\section{Introduction}
\label{sec:intro}

With the continuous development of 5G technologies, 5G-related services inevitably have begun to enter indoor environments, with signal coverage provided by microcells or femtocells. The deployment location, power setting, resource allocation, and antenna gain of such small cells will greatly affect the quality of service (QoS) for UEs. Therefore, there is rich research targeting the optimization of small-cell-related settings. Recently, machine learning (ML)-based, especially reinforcement learning (RL)-based algorithms appear attractive in this domain because of their proven success in solving complex optimisation problems. For instance, the interference control in a heterogeneous network utilising Q-learning was discussed in \cite{bennis2010q}. Similar Q-learning-based power control for indoor voice over LTE radio bearer was proposed by \cite{mismar2018q}. Recently, Mismar et~al.\ put forward a deep Q network (DQN)-based method for joint beamforming, power control, and interference coordination \cite{mismar2019deep}. \PZ{In \cite{amiri2019reinforcement}, multi-agent RL (MARL) is adopted to realise self-organising and power control in heterogeneous networks. In comparison, the MARL method is adopted in \cite{wang2020decentralized} to tackle interference mitigation for indoor coverage for 5G (and beyond) systems.
In \cite{neto2021uplink}, the authors put forward an RL framework for uplink power control. A federated DQN approach for user access control is proposed in \cite {cao2021federated} under the context of O-RAN.}

However, it is noticeable that existing approaches are discussed without differentiating scenarios, and are mainly for outdoor macro cells. The underlying assumption is that such environments share the common signal transmission proprieties and UE patterns, so that the trained RL models can be applied to other scenarios. However, such assumptions do not hold for indoor scenarios. As the UE moving patterns largely depend on (or are limited by) the layout of the room, the optimal model in one room can be drastically different from others, i.e.\ the model is room-dependent, which is difficult to serve in other rooms. In order to increase the model's generalisation ability, a training process considering multiple indoor environments is needed. Also, from the view of the indoor network provider, it is necessary to have a general model that can be deployed into a new scenario with zero or minimal amount of learning. Meanwhile, the training process ought to be controllable and not consume too much bandwidth.

The aforementioned two considerations motivate us to develop a  federated reinforcement learning (FRL) framework in this paper. \PZ{The adoption of FRL involves the updates of RAN at the hardware and cloud system and correspondingly, the data collection and model deployment pipeline. Fundamentally, the O-RAN architecture enables the feasibility of executing the ML/RL model through radio intelligence controllers (RICs) \cite{li2021rlops}.}
\PZ{For each room, an independent RL agent is needed, while all rooms together cerate the federated learning (FL) learning paradigm. It is worth pointing out that the definition of ``state'' in the RL model used here relies only on off-shelf cell information, like CQI.} FRL is promising because it involves neural network parameters communication rather than real user data, which removes privacy concerns for indoor UE information. Meanwhile, the global model trained by FRL is able to adapt to a new environment more rapidly. The FRL framework is a step towards intelligent RAN. 
The contributions of this paper are summarised below:

\begin{itemize}
\item For indoor cell transmit power control, this is the first work that considers the variation of RL model training in different room layouts.

\item We put forward an FRL framework to solve the  generalisation, distribution, and adaption problems of the model under the context of O-RAN.

\item Extensive simulations are performed to demonstrate the gains on throughput and generalisation ability of the proposed method.

\item The simulation process strictly follows the hierarchical orchestration structure of O-RAN, where the RICs are established on top of the simulation environment. It provides a simulator design paradigm compatible with ML and RL. \PZ{The document of code for the entire simulator is available at:  https://aimm.celticnext.eu/simulator/}.

\end{itemize}

\section{Background}
\label{sec:background}
\subsection{O-RAN}
O-RAN is a new emerging architecture for the radio access network. It is attracting much attention due to two proprieties: openness, and intelligence. Openness means that it adopts a standard and well-defined hardware interfaces and software services, so the equipment or IPs involved in O-RAN is not vendor-specific. More importantly, it embraces artificial intelligence (AI) in its basic standard formulation. Two types of RIC are designed in O-RAN to realise intelligent control of the entire network: non-real-time RIC and near-real-time RIC. AI or ML models can be deployed into RICs in the form of microservice applications, i.e.\ xApps and rApps.

\subsection{RL and DQN}
\label{subsec: RL and DQN}
RL is a class of learning paradigms in ML. The agent focuses on the actions of interacting with the environment in order to achieve the largest accumulative rewards. DQN is a relatively mature and a widely used algorithm of RL. It has been proposed for controlling complex video games only using images \cite{mnih2015humanlevelcontrolDQN}. The idea of DQN lies in the use of deep neural networks \eqr{f_\theta}, to estimate the \PZ{state ($s$)-action ($a$) value ($Q$ value)}, that it \eqr{f_\theta = Q(s,a)}. Taking the \PZ{action $a$} in the given space, the optimal policy can be constructed as:
\begin{equation}
    \pi^*(s) = \arg \max Q^*(s,a).
\end{equation}
\eqr{Q^*(s, a)} obeys the Bellman optimality equation \cite{sutton2018reinforcement}:
\begin{equation}
\label{eq:bellman optimality}
    Q^\pi(s,a) = \mathds{E}_{s^\prime \sim \mathcal{S}} \Big[ r + \gamma \max_{a^\prime} Q^* (s^\prime, a^\prime) \big| s,a\Big].
\end{equation}
To learn the $Q$ value at iteration \eqr{i}, the following loss is minimised with respect to \eqr{\theta}:
\begin{equation}
    L_i (\theta_i) = \mathds{E}_{s,a \sim \rho(\cdot)} \Big[ (y_i - Q(s,a;\theta_i))^2 \Big],
\end{equation}
where
\begin{equation}
    y_{i, Q} = \mathds{E}_{s^\prime \sim \mathcal{S}} \Big[ r + \gamma \max_{a^\prime} Q(s^\prime, a^\prime; \theta_{i-1} \big| s,a) \Big].
\end{equation}
Meanwhile, an experience replay mechanism and a target network are introduced in DQN to stabilise the training process.

\subsection{FL}
\label{subsec: FL}
FL is an ML setting in which multiple clients collaboratively train a model under the orchestration of a central server, while keeping the training data decentralized \cite{kairouz2021advances}. Due to concerns of privacy and communication efficiency, the training paradigm is that local models need to upload the model parameters to the global model (in the central server), and the global model returns the model parameters after parameters aggregation. In this paper, we apply FL to the RL paradigm. 

\subsection{Simulator Design}
\label{subsec: simulator design}
In this paper, the simulator not only plays the role of radio link simulations, but also undertakes the work of office layout and UE trajectory generation. It is also the venue of RL agent and FL instantiation and training. Meanwhile, the trained model aims to be transferred to the O-RAN. Hence, the interfaces between RL agents and radio simulations should be specified. The definition of the simulator's functionalities obeys the hierarchical architecture of O-RAN strictly. It adopts a process-based discrete-event simulation framework, so different processes like \textit{RICs}, \textit{Logs}, and \textit{mobility management entities (MMEs)} execute in parallel without interfering with each other. 

\section{Problem Formulation}
\begin{figure}[t]   
        \centering   
        \includegraphics[width=0.8\columnwidth]{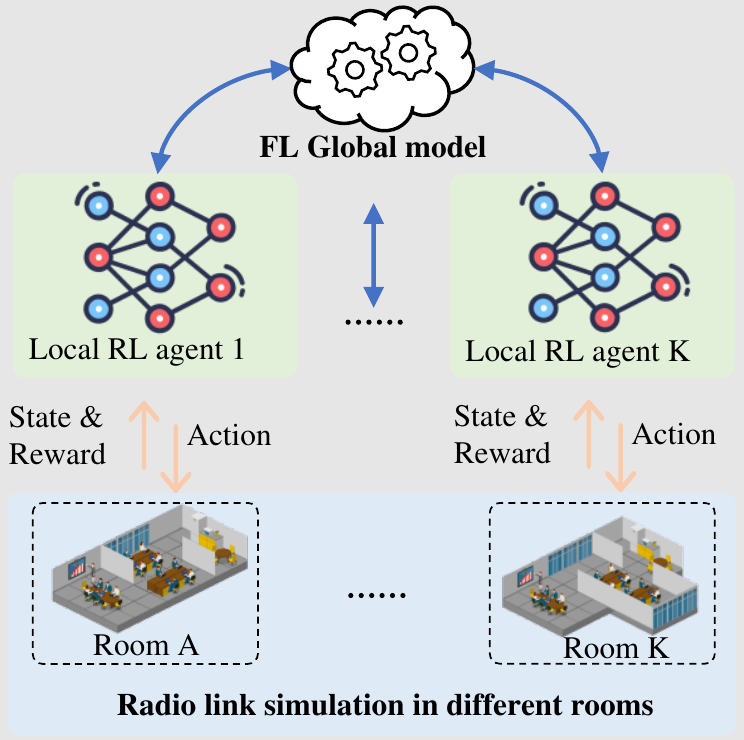}   
            \caption{System diagram of FRL framework} \label{fig:system figure}
\end{figure}

This paper studies a scheme for transmission power control of small cells in distributed indoor environments. Each indoor environment is viewed as being controlled by an independent local RL agent, while multiple RL agents are orchestrated by FL. FL distributes the model from the central server to the local agent, and also aggregates local models to a new global model periodically. The system diagram is shown in Fig.~\ref{fig:system figure}.

\subsection{Cells Transmit Power Control in a Single Room}
In this paper, it is assumed that there are $M$ cells and $N$ UEs in a single room and no subband or physical resource block allocation is considered. The downlink data rate $C_{m,n}$ from the cell $m$ to UE $n$ can be modelled as follows:
\begin{equation}
    C_{m,n} = B_m \log_2 \left(1 + \mathrm{SINR}_{m,n}\right),
    \label{eq:C}
\end{equation}
where $B_m$ is the bandwidth of the cell $m$ and $\mathrm{SINR}_{m,n}$ is the Signal-to-Noise plus Interference Ratio (SINR), which is determined for the transmission from cells to UEs.  
The $\mathrm{SINR}_{m,n}$ is defined as follows:
\begin{equation}
    \mathrm{SINR}_{m,n} = \frac{G_m G_n P_m \ell^{(m)}(d_n)}{W_m+\sum_{i=0,i\neq m}^{M} I_{i,m}}
    \label{eq: SINR}
\end{equation} 
where
\begin{itemize}
    \item $G_m$, $G_m$ are the transmission and receiver antenna gains.
    \item $P_m$ signifies the transmission power of cell $m$.
    \item $\ell^{(m)}(d_n)$ expresses the path-loss at a distance $d_n$ (between cell $m$ and UE $n$). 
    \item $W_m$ represents the thermal noise power.
    \item $I_{i,m}$ is the interference power received.
\end{itemize}
In the process of code implementation, the $\mathrm{SINR}_{m,n}$ will be first converted to the corresponding Channel Quality Indicator (CQI) value, then the final data-rate is calculated according to the relationships demonstrated in Table~\ref{table:cqi}.

\begin{table}[t]
\centering
\caption{CQI Table \cite{eskandari2022smart}}
\label{table:cqi}
\begin{tabular}{cccc} 
\toprule
SNR(dB) & CQI Index & Modulation & Code Rate($\times$ 1024)\\ \midrule
$-\infty$ & 0 & Out of Range & \textemdash \\ \midrule
-6.9360  & 1 & QPSK & 78 \\ \midrule
-5.1470  & 2 & QPSK & 120 \\ \midrule
-3.1800  & 3 & QPSK & 193 \\ \midrule
-1.2530  & 4 & QPSK & 308 \\ \midrule
0.7610  & 5 & QPSK & 449 \\ \midrule
2.6990  & 6 & QPSK & 602 \\ \midrule
4.6940  & 7 & 16-QAM & 378 \\ \midrule
6.5250  & 8 & 16-QAM & 490 \\ \midrule
8.5730  & 9 & 16-QAM & 616 \\ \midrule
10.3660  & 10 & 64-QAM & 466 \\ \midrule
12.2890  & 11 & 64-QAM & 567 \\ \midrule
14.1730  & 12 & 64-QAM & 666 \\ \midrule
15.8880  & 13 & 64-QAM & 772 \\ \midrule
17.8140  & 14 & 64-QAM & 873 \\ \midrule
19.8290  & 15 & 64-QAM & 948 \\ 
\bottomrule
\end{tabular}
\end{table}

\textbf{Optimisation Objective:} For a local RL agent, the optimisation objective is to maximize the overall throughput of the entire room. The objective function is written as:
\begin{equation}
\begin{split}
    \max \sum_{m\in M} \sum_{n \in N} C_{m,n},  \\ \text{s.t. } P_m \in P_\text{POT},
\end{split}
\label{eq:objective}
\end{equation}
where $P_\text{POT}$ is the set of possible transmission power levels.
\begin{algorithm}[t]
\SetAlgoLined
\LinesNumbered
\SetKwProg{Function}{function}{}{end}
\SetKwRepeat{Do}{do}{until}
\textbf{Require:} The setting of different indoor scenarios.

Initialize \eqr{K} clients with network \eqr{Q_k} and \eqr{\hat{Q_k}}, and Global model \eqr{Q_\text{Global}}.

Initialize the experience replay memory \eqr{D}.

Initialize the agent to interact with the environment \eqr{E_k}.




\While{not Done}{
\For{\eqr{k=1,K}}{  

Update model \eqr{Q_k} by \eqr{Q_\text{Global}}
 
\For{\eqr{t=1,T}}{
Reset the environment

Set the initial state \eqr{s = s_0}

With probability \eqr{\epsilon} select a random action \eqr{a_t}

Otherwise \eqr{a_t = \arg \max_a Q_k(s_t, a; \theta, W)}

Execute action \eqr{a_t} in environment $k$ and observe reward \eqr{r_t} and new state \eqr{s_{t+1}}

Store transition \eqr{(\phi_t, a_t, r_t, \phi_{t+1})} in \eqr{D}

Sample random minibatch of transitions \eqr{s_j, a_j, r_j, s_{j+1})} from \eqr{D}


Perform a gradient descent step on \eqr{(y_{i} - Q_k(\phi_j, a_j; \theta, W))^2}

For every \eqr{C} steps, set \eqr{\hat{Q_k} = Q_k}

}

Upload model \eqr{Q_k} to \eqr{Q_\text{Global}}

Wipe \eqr{D}

}

For every $E$ cycles, aggregrate the global model \eqr{Q_\text{Global}} using equation \ref{eq:FedAvg}.

}

\caption{Federated DQN for power adjustment of indoor cells.}
\label{al:FRL}
\end{algorithm}
\subsection{Markov Decision Process}
We formulate the problem in (\ref{eq:objective}) as a finite Markov decision process (MDP).
An MDP is defined by the tuple $(\mathcal{S}, \mathcal{A}, \mathcal{P}, \mathcal{R}, \gamma)$, where the set of environment states is represented by $\mathcal{S}$; $\mathcal{A}$ is the action space of agent; $\mathcal{P}$ is the transition probability from state $s\in \mathcal{S}$ to state $s^\prime \in \mathcal{S}$ for any given action $a \in \mathcal{A}$, and $\mathcal{R}$ is the reward function. $\mathcal{\gamma}$ is the discount factor. 

\textbf{Steps and Episodes:} 
In one room, an episodic task is defined. Each episode contains 100 sequential steps, while UEs move to new locations at each step. 
The possible locations of UEs are initially pre-generated by a ``billiard''  model \cite{briggs_1988}, when initialising the room layout. In this model, users bounce off walls. In the sequential decision-making problem, the RL agent looks for the optimal combination of transmission powers for all cells at each step and naturally in every episode. At each step, when a new action is performed, handover events for all UEs will be triggered immediately. The handover decision is based on Reference Signal Received Power (RSRP), which means that the UE always attaches to the cell with the highest RSRP.

Furthermore, at time $t$, for $m\in M$ and $n\in N$, the state, action and reward of the deep RL agent are defined as below:
\begin{itemize}[leftmargin=*]
    \item \textbf{State:} The state $s_t$ is consist of three parts: 1) the normalised transmission power of current $M$ cells $P_{m,t}$; 2) the number of UEs attached in the each cell $NU_{m,t}$; 3) the CQI information of all UEs reported to different cells, that is $CQI_{m,n,t}$. So the overall state is:
\begin{equation}
    s_t = (P_{m,t}, NU_{m,t}, CQI_{m,n,t})
\end{equation}
    \item \textbf{Action:} The action $a_t$ at time $t$ is to select a transmission power for each cell. The action space is discrete. 
\item\textbf{Reward:} The training criterion is the throughput of all UEs. In this paper, under a joint consideration between the maximum throughput and the QoS guarantee for UEs, we take the 0.25 lower quantile of the distribution of throughput across all UEs, denoted  $\mathbf{Q_1}$, as the optimisation objective. So the reward $r(s_t,a_t)$ of executing action $a_t$ at state $s_t$ is defined as the quantile improvement of the entire network after this action, and any subsequent handover, have taken effect.
\begin{equation}
    r(s_t,a_t) = \mathbf{Q_1}(a_t) - \mathbf{Q_1}(a_{t-1})
\label{eq:reward}
\end{equation}
It should be noticed that the reward design for this optimisation problem is flexible and goal-related. It depends on the focus of optimisation. For instance, we formalise the reward to equation~(\ref{eq:reward}) because we are concerned about the lowest QoS guarantee for all UEs. However, if more attention is paid to the balance of workload of cells, the reward can easily be redefined.
\end{itemize}

\subsection{The FRL Algorithm}
\PZ{FRL is a promising and efficient method of RL to create a distributed paradigm and so preserve data privacy. In our case, FRL consists of multiple independent RL agents serving multiple rooms. Each local agent is acting as the cell's transmit power controller for one room based on the global DQN model, which is aggregated by an FL algorithm, as shown in Fig.~\ref{fig:system figure}. We adopt \textit{Fedavg} as the default algorithm for global model aggregation \cite{mcmahan2017communication}}. For local agent $k\in K$ with model parameters $\theta_k$, the aggregation operation is expressed by equation~(\ref{eq:FedAvg}):
\begin{equation}
\PZ{\theta_\text{Global} = \frac{1}{K}\sum_{k=0}^{K}\theta_{k}}.
\label{eq:FedAvg}
\end{equation}

Local agents upload their model parameters to the central server every $E$ cycles. Then the global model will be broadcast back to all agents after aggregation, and the global model serves as the pre-trained model for each agent after broadcasting. \PZ{The RL models will be installed in the RICs of O-RAN through the form of xApps or rApps, to perform local training and parameter uploading, while the global model can be deployed in the central server of network operators.} The overall FRL training scheme is illustrated in Algorithm~\ref{al:FRL}.

\begin{figure*}[t]   
    \subfloat[\label{fig:layout_room1}]{
      \begin{minipage}[t]{0.5\linewidth} 
        \includegraphics[width=2.4in]{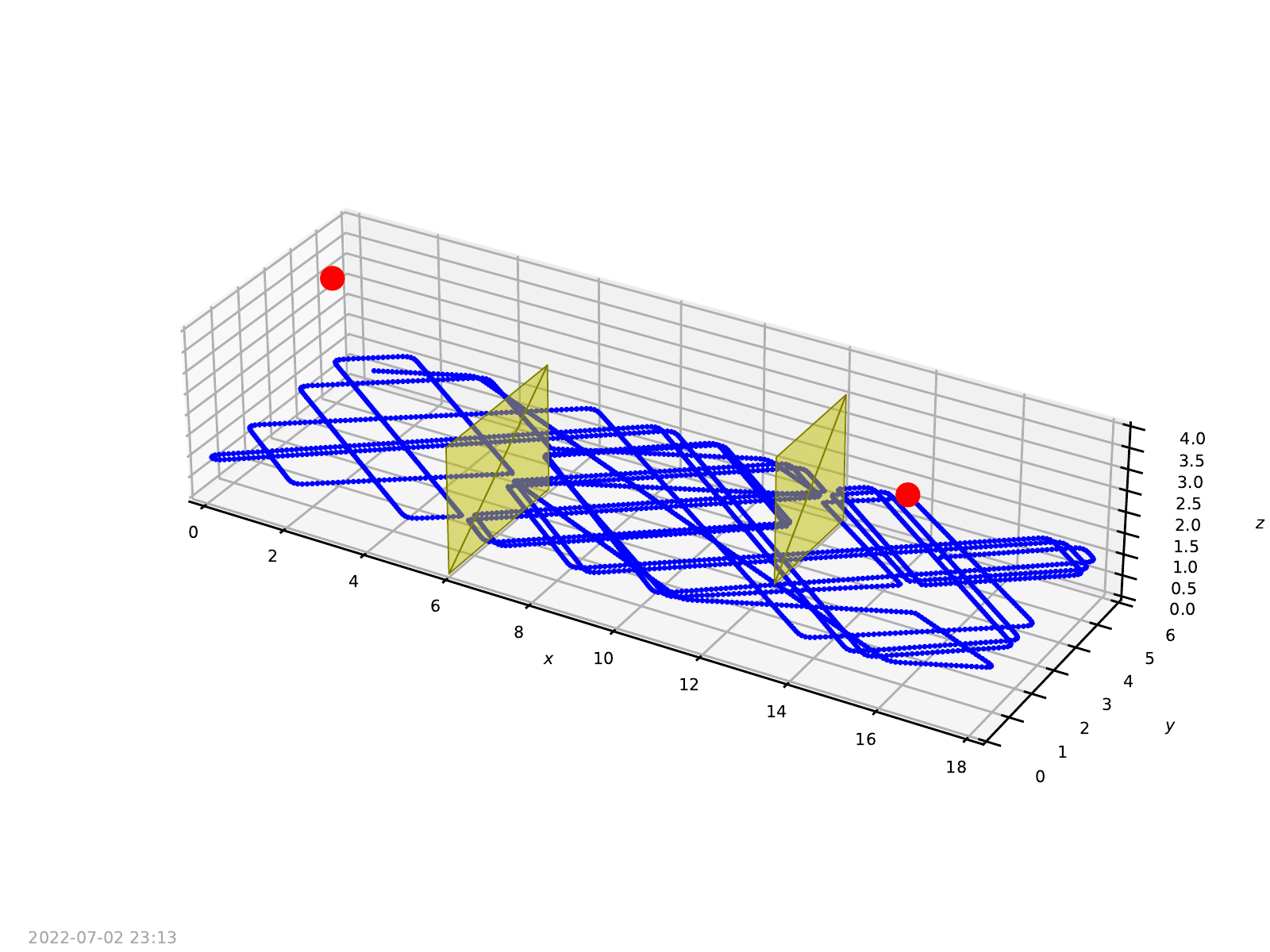}   
      \end{minipage}%
      }
      \hfill
        \subfloat[\label{fig:layout_room2}]{
      \begin{minipage}[t]{0.5\linewidth}   
        \centering   
        \includegraphics[width=2.4in]{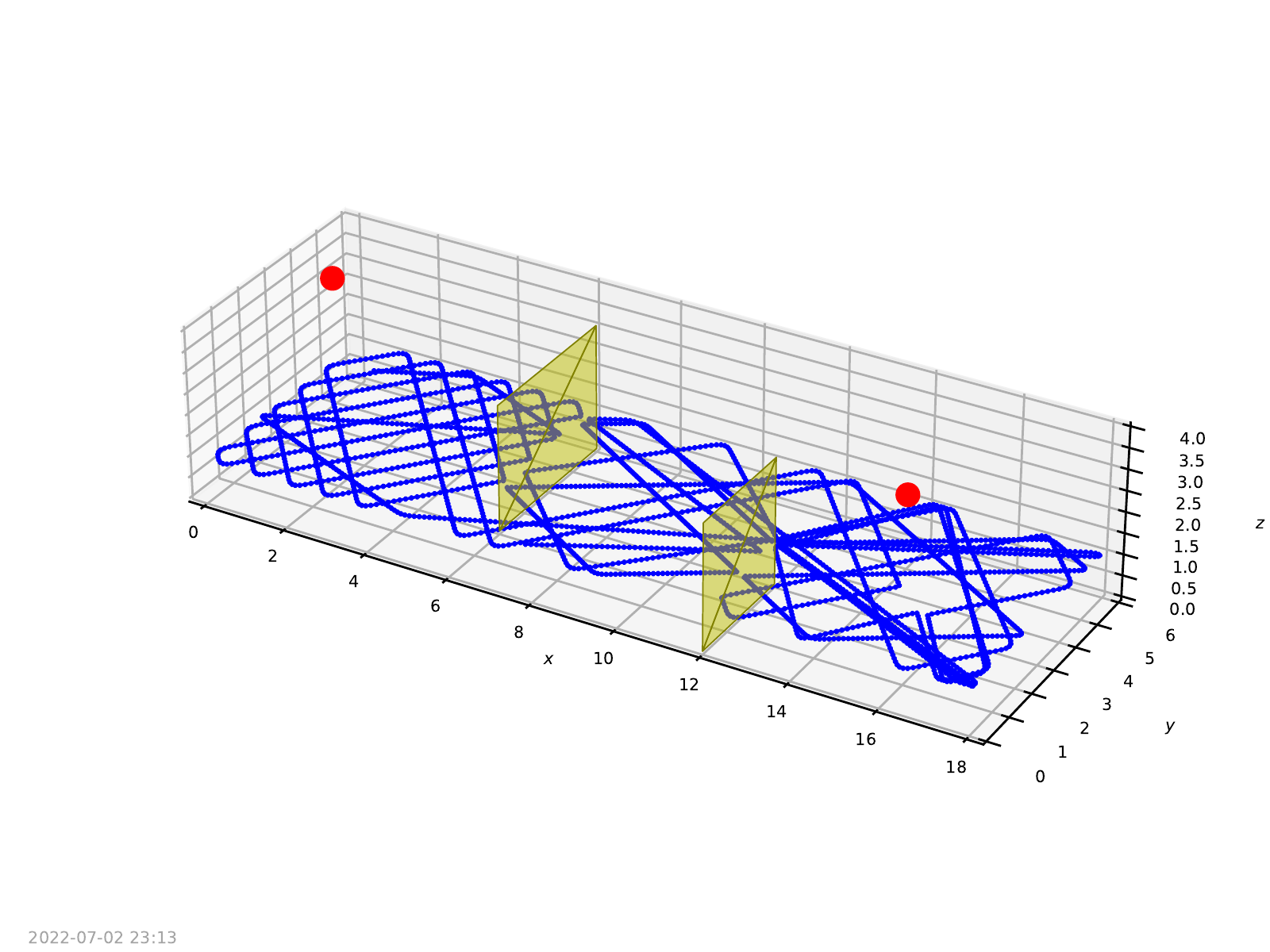}   
      \end{minipage} 
      }
     \\ \vspace{0.1cm}
     \subfloat[\label{fig:layout_room3}]{
      \begin{minipage}[t]{0.315\linewidth}   
        \centering   
        \includegraphics[width=2.1in]{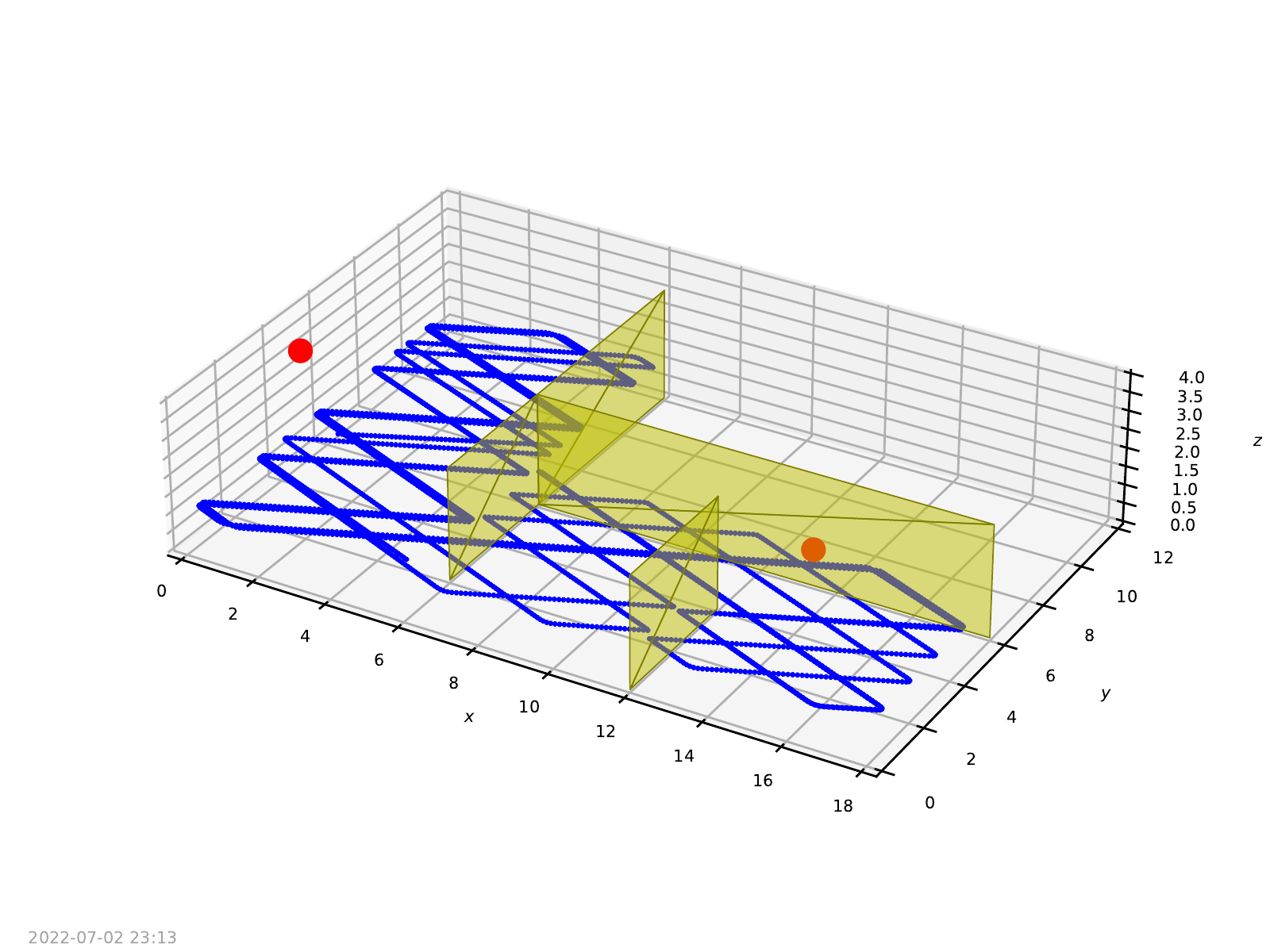}   
      \end{minipage}  
      }
    \subfloat[\label{fig:layout_room4}]{
      \begin{minipage}[t]{0.315\linewidth}   
        \centering   
        \includegraphics[width=2.1in]{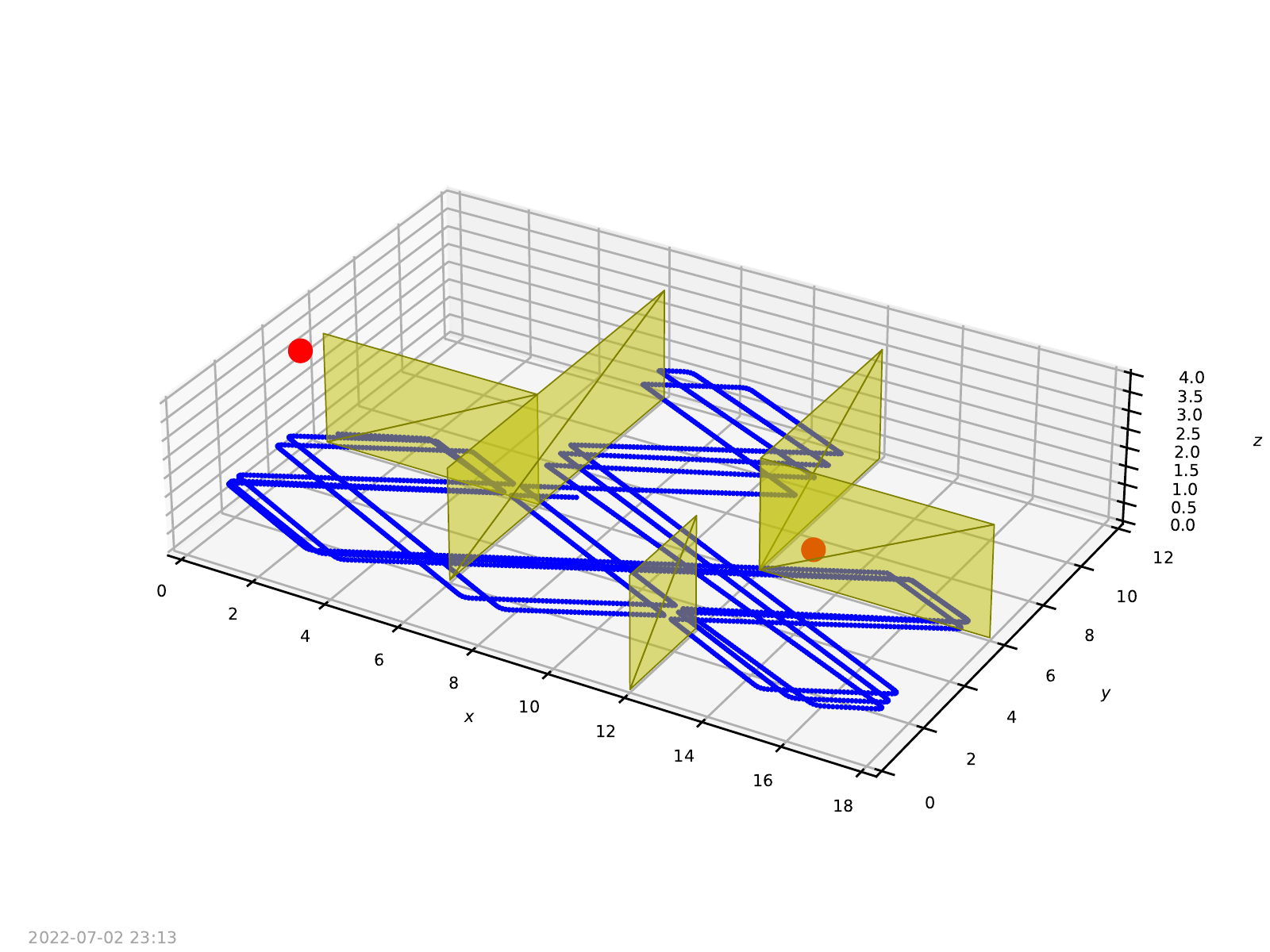}   
      \end{minipage}  
      }
     \subfloat[\label{fig:layout_room5}]{
      \begin{minipage}[t]{0.315\linewidth}   
        \centering   
        \includegraphics[width=2.1in]{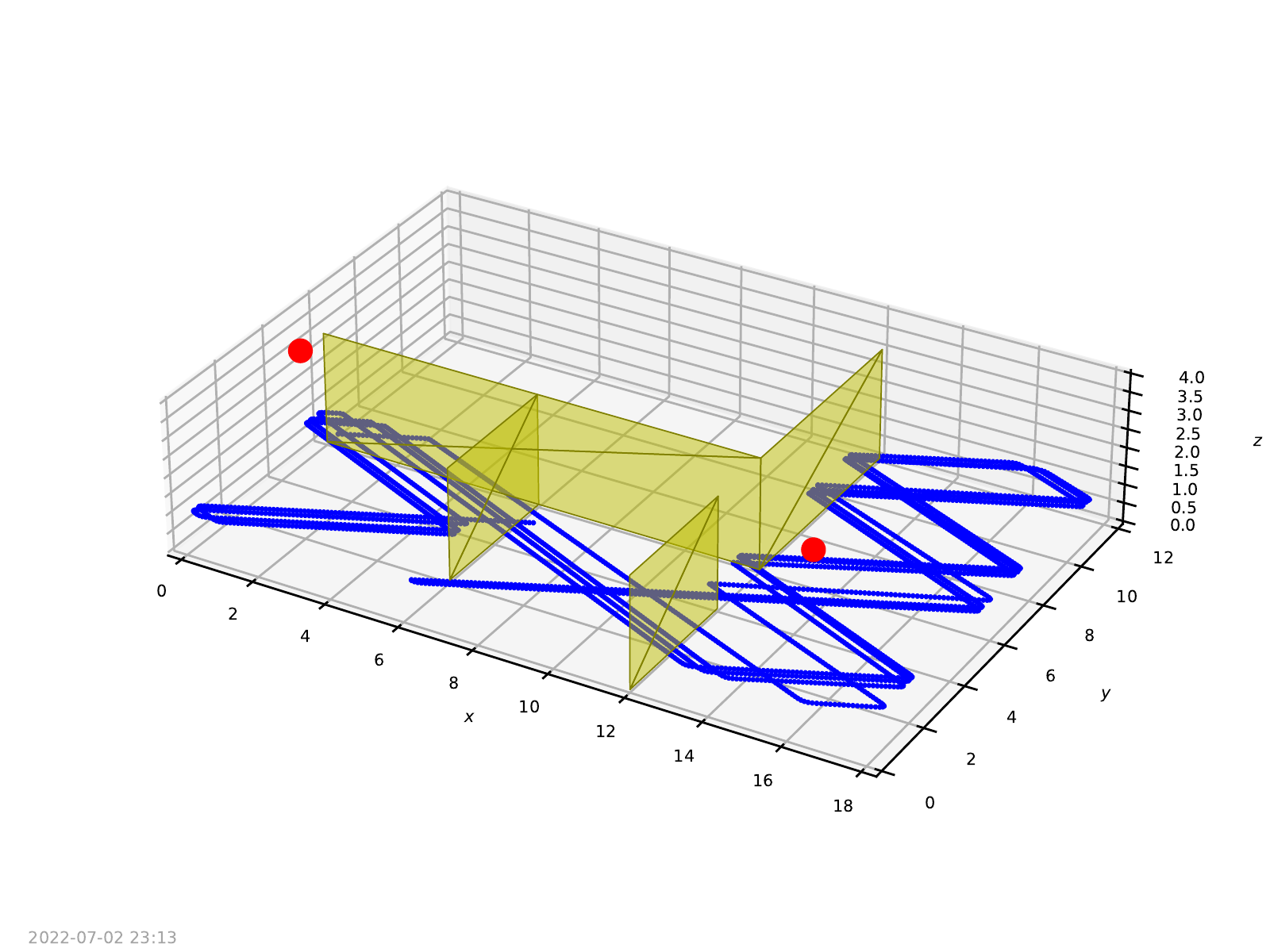}   
      \end{minipage}  
      }
      \caption{(a)--(e) show different 3D layouts of rooms (room A--E). The yellow panels are the interior walls. The blue lines are possible UE locations generated by the billiard algorithm. Red dots indicate the locations of cells} \label{fig:layout_rooms}
    \end{figure*} 

\section{Simulation}
\label{sec:simulation}
\subsection{Room Layouts and User Mobility Mode}
To evaluate the performance of the proposed FRL scheme, we defined five typical room layouts. As shown in Fig.~\ref{fig:layout_rooms}, all rooms are of 4m height. Rooms A and B are narrow rectangular layouts with a size of $18\times6$m. Room C is L-shaped and room D is T-shaped, while room E is L-shaped in another direction. They are all of dimension $18\times12$m. The yellow rectangles represent the interior wall panels of the room; the red points are the indoor cells deployed. \PZ{The blue traces are the potential UEs locations generated by the Billiard model \cite{briggs_1988}. Since we are dealing with heterogeneous UE distributions, which are directly caused by different room layouts, the billiard model can reflect the room layout information as much as possible, which is helpful in evaluating the RL performance. For the same reason, we maintain the positions of edge users that seem to penetrate the room's interior walls.} The user trajectories are sampled from these locations according to the number of steps in each episode. To increase the generalisation ability of the RL agent, at each step, we add random position offsets in initial $x,y$ locations respectively, which are sampled from Gaussian distributions with mean $\mu=0$ and variance $\sigma=0.5$. The height of all UEs is fixed at $1$m.

\subsection{Radio Simulation Setting}
\label{subsec:radio setting}
The cells modelled in this paper follow the 5G gNodeB architecture. \PZ{According to engineering experience, we reasonably assume that $M=2$ cells and $N=30$ UEs are typical for one large room. The five rooms are initialised for FRL to reduce the simulation complexity.} For parameters related to Eqs.~(\ref{eq:C}--\ref{eq: SINR}), $B_m$ is 20 MHz; $G_m$ and $G_n$ are $0$ dBi; $W_m$ is constant. The initial transmission power is $24$ dBm. The indoor path-loss model shown in equation~(\ref{eq:pathloss model}) is used for $\ell$, which comes from 3GPP TR 36.873 version 12.7.0 Release 12.
\begin{equation}
\begin{split}
& \ell_\text{Los} = 22\log_{10}{d_{3D}}+28.0+20\log_{10}{f_c} \\
& \ell_\text{NLos} = 36.7\log_{10}{d_{3D}}+22.7+26\log_{10}{f_c}-0.3(h_\text{uT}-1.5),
\end{split}
\label{eq:pathloss model}
\end{equation}
where ${f_c}$ is $3.5$ GHz, and $h_\text{uT}=1m$.
It is to be noted that 3GPP updates the indoor propagation models in different releases, but the variation between such models is minor and has negligible influence on our RL training. 

For every cell, we assume $P_\text{POT.}=[19.5, 21.0, 22.5, 24.0]$ dBm, i.e.\ there are four power levels for each cell. Considering the total number of power levels for $2$ cells, there are 16 combinations. After excluding those combinations with the same interference proportion, the action-space size of the RL agent is $11$.

\subsection{FRL Setting}
\label{subsec:FRL setting}
DQN and FedAvge are adopted as algorithms for the proposed FRL framework. $Q_k$ and $\hat{Q_k}$ are deep neural networks with fully-connected layers; hyperparameter details can be found in Table~\ref{table:Hyperparameters}. 
The hyperparameters in this table are reasonable empirical values, based on our experience of running many simulations with varying values.
The RL agents of the first four rooms (A--D) are used for the federated global model training, and room~E is used for the validation of the FRL model.

\subsection{Baseline: exhaustive search}
\label{subsec:Heuristic}
To provide a reliable baseline for evaluation of the FRL method,  \PZ{we exhaustively search through all allowed power levels and then select the power setting which achieves the highest throughput. This is guaranteed to correctly maximize equation~(\ref{eq:objective}), and is feasible in our test scenarios because of the small problem size.}

\begin{table}[t]
\centering
\caption{Hyperparameters of Federated Reinforcement Learning}
\label{table:Hyperparameters}
\begin{tabular}{ll} 
\toprule
Name & Value\\ \midrule
FL algorithm & FedAvg \\ \midrule 
Number of Clients K & 5 \\ \midrule 
Aggregation cycles E & 380 \\ \midrule 
RL algorithm & DQN \\ \midrule
Exploration rate $\epsilon$ & 0.9\\ \midrule
Batch size & 128\\ \midrule
Maximum timesteps in each episode & 100 \\ \midrule
Target network update interval & 100 \\ \midrule
Reward discount factor$\gamma$ & 0.98 \\ \midrule
Optimizer & Adam \\ \midrule
Learning rate & 0.001 \\ \midrule
Layer type & fully connected layer \\ \midrule
Number of neurons of each layer & [200,100,50] \\ \midrule
Activate function (not for output layer) & Relu \\ \midrule
Activate function for output layer & Linear \\ 
\bottomrule
\end{tabular}
\end{table}

\section{Results}
\label{sec:results}
\subsection{Training of the Single RL Agent and FRL}
\begin{figure}[t]   
        \centering   
        \includegraphics[width=0.85\columnwidth]{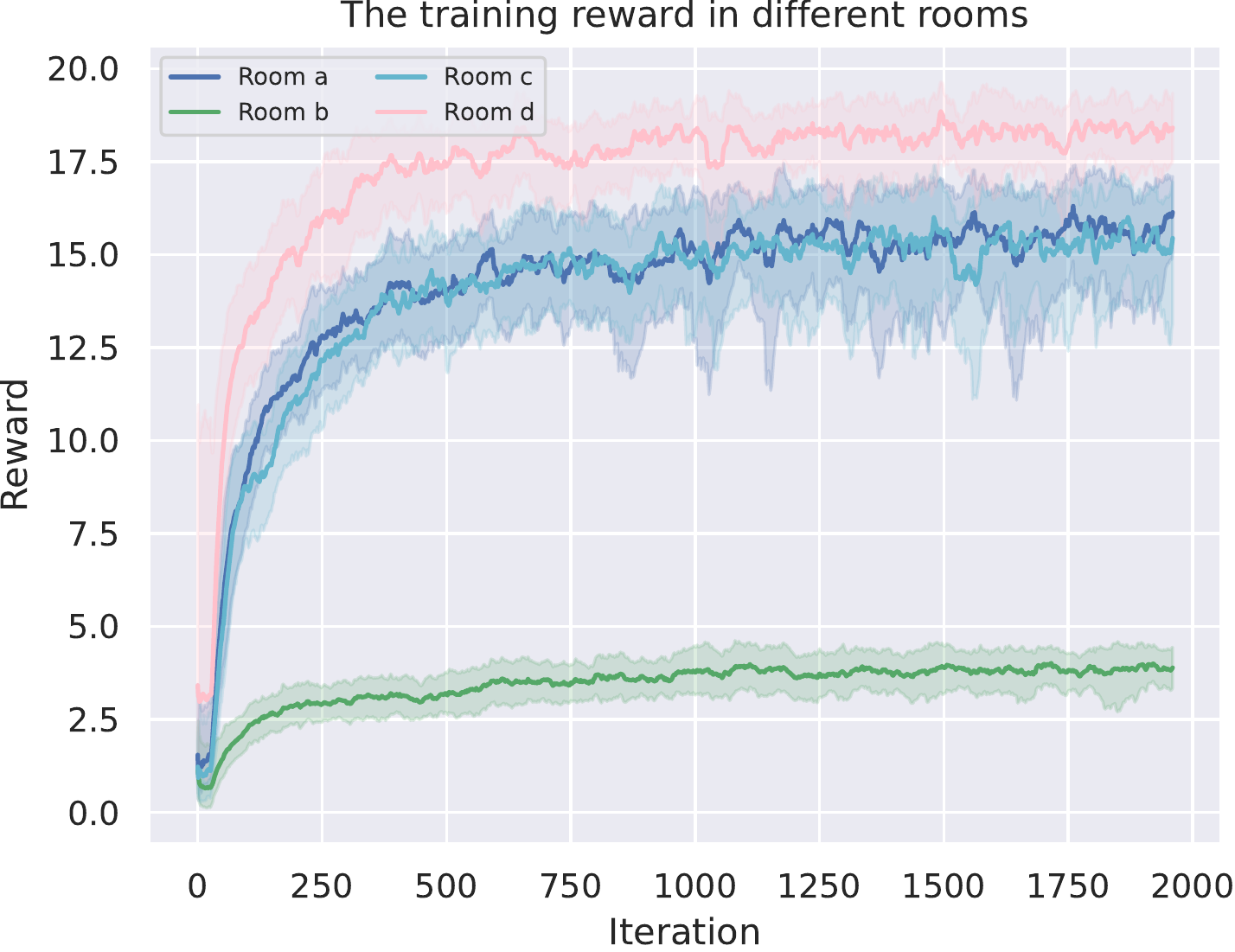} 
            \caption{RL training reward in different rooms} \label{fig:rewards}
\end{figure}

The reward during training RL agents in room A-D are illustrated in Fig.~\ref{fig:rewards}, where each agent is trained independently five times, to evaluate the amount of variantion in the training process. Each training phase lasts around $2000$ episodes.
It can be seen that the single RL agent works well for the corresponding scenario. Although the convergence time varies, a stable reward gain can always be observed. It is noticeable that the reward varies in each room; this is because of the heterogeneity of the UE distribution across different rooms.

\begin{figure}[t]   
        \centering   
        \includegraphics[width=1\columnwidth]{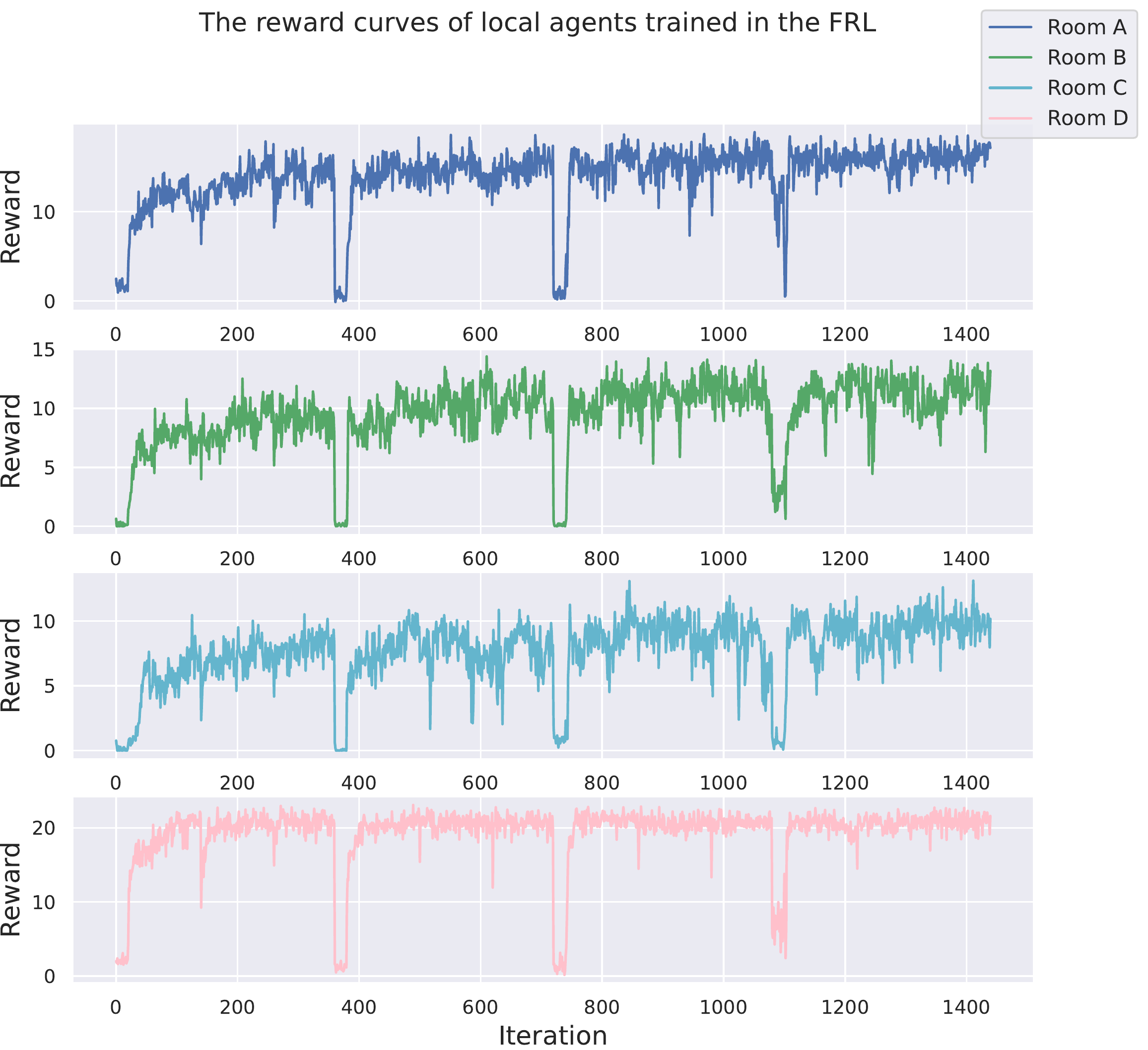}   
            \caption{The reward of agent trained in the FRL} \label{fig:FL}
\end{figure}

As discussed in the above section, we train an FRL global model using rooms A--D. The global model aggregation happens every $E=380$ cycles. The training curves of local clients (A--D) are demonstrated in Fig.~\ref{fig:FL}. It can be observed that the reward drops every 380 cycles; this is where the aggregations happens. The whole FRL training process ends with the convergence of each client. 

In the single RL validation stage, the trained model is frozen and deployed in the same environment as in the training stage. We calculate the cumulative throughput of the entire network based on the 0.25 quantile and average data-rate of all episodes. \PZ{The results of the random power allocation, RL model and exhaustive search method are shown in Table~\ref{table:tp comparsion}. It can be seen that the RL algorithms outperform both the random allocation and exhaustive search method in any environment, and the trained global model of FRL shows a similar performance compared with signal FL.
Moreover, it is noticeable that the trained RL model shows great advantages in terms of inference time. When the UE locations change, the DQN only needs to make one forward inference to get the optimal transmit power setting, which is a capability that the greedy algorithm can't match.} 

\begin{table}[t]
\centering
\caption{The cumulative throughput compare based on 0.25 quantile ($Q_1$) and average for all UEs (in Mbps)}
\label{table:tp comparsion}
\begin{tabular}{llrrrr} 
\toprule
Algorithm                 & Criterion & Room A  & Room B  & Room C  & Room D   \\ \midrule
\multirow{2}{*}{\PZ{Random}} & $Q_1$       & 93 & 66 & 62 & 46  \\ 
                           & Avg.   & 122 & 109 & 117 & 65  \\ \midrule
\multirow{2}{*}{Exhaustive} & $Q_1$       & 103 & 156 & 109 & 112  \\ 
                           & Avg.   & 179 & 207 & 216 & 184  \\ \midrule
\multirow{2}{*}{Single RL}  & $Q_1$      & 114 & 163  & 118 & 113   \\ 
                           & Avg.   & 221 & 223 & 218 & 225  \\\midrule
\multirow{2}{*}{FRL}  & $Q_1$      & 115 & 164  & 113 & 112   \\ 
                           & Avg.   & 219 & 223 & 219 & 225	  \\
\bottomrule
\end{tabular}
\end{table}


\subsection{Adaption Test of FRL Global Model}
\begin{figure}[t]   
        \centering   
        \includegraphics[width=0.85\columnwidth]{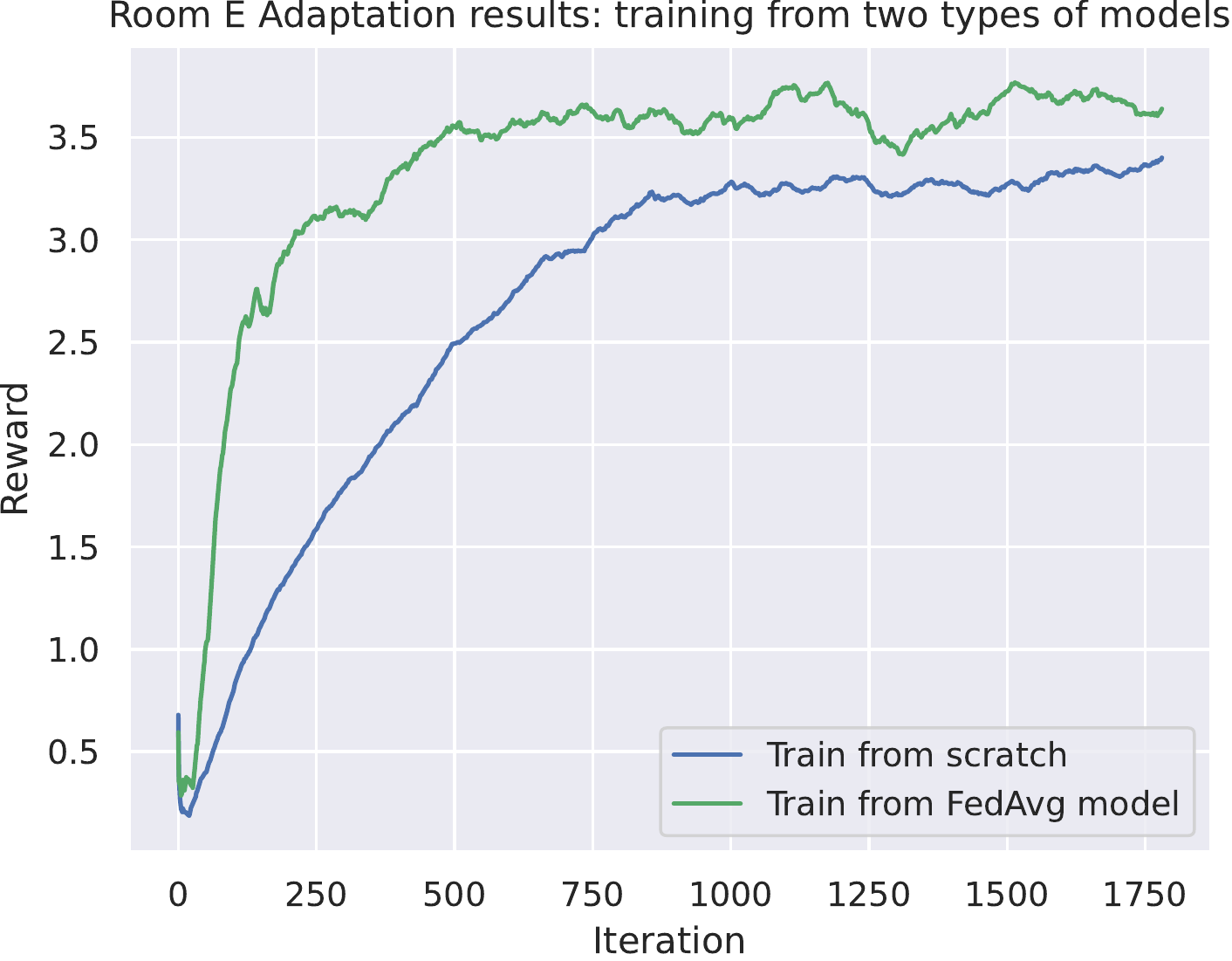}   
            \caption{Room C's RL model training from different pre-trained models} \label{fig:adaptation}
\end{figure}

To validate the generalisation and adaptation ability of the FRL approach, the model trained in rooms A--D is tested in a new environment (room E). Two single RL agents are trained. One is trained from the scratch, another one is trained from a FRL model pre-trained in room A--D. The comparison can be found in Fig.~\ref{fig:adaptation}. It is obvious that the adaptation of pre-trained FedAvg global shows significant advantages in training speed and final performance. The RL agent trained from the FedAvg model converges faster than all others. This reveals that the knowledge learned in the global model can guide the model training in a new environment. 

\section{Discussion}
\label{sec:discussion}
We utilised FRL to solve indoor small cell transmission power control problem. The FRL framework ensures the privacy and security of UEs and is able to provide a template for model distribution, which fits the xApps model framework. Network operators may moved towards intelligent networks with our proposed methods. For the RL-based controller, there are still some problems waiting to be explored. One of the problems is that when we increase the number of cells or add other optimization options, the action space will grow exponentially, which can lead to a large increase in training costs. Although some schemes such as action-space encoding and actor-critic structure can partially solve this problem, the effect is not satisfactory. On the other hand, we assume the path-loss models in the different indoor environments are the same, but in reality, due to the multipath effect of indoor environments, such empirical models are not reliable, which results in the simulation-versus-reality issue needing to be addressed \cite{li2022sim2real}. \PZ{In the future, we will consider the joint optimisation of transmit power, physical resource block, loading balance etc.; all these optimisations will be unified in our proposed FRL approach.}

\section{Conclusions}
\label{sec:Conclusion}
This paper discusses the issue of indoor cell transmit power control in the context of O-RAN, emphasizing the room-dependent properties and lack of generalisation ability of a single RL model. Based on this, we propose an FRL framework. The client is in a single indoor environment and learns the best policy by RL. All clients will periodically upload model parameters and integrate them in the global model. The global model will act as the base model for learning in new environments. The simulation results demonstrate the feasibility and advantages of the proposed method, both in throughput and the learning efficiency.

\section*{Acknowledgment}
This work was developed within the Innovate UK/CELTIC-NEXT European collaborative project on AIMM (AI-enabled Massive MIMO). This work has also been funded in part by the Next-Generation Converged Digital Infrastructure (NG-CDI) Project, supported by BT and Engineering and Physical Sciences Research Council (EPSRC), Grant ref. EP/R004935/1.
\bibliographystyle{IEEEtran} %
\bibliography{IEEEabrv,references} 
\end{document}